\documentclass{PoS}

\usepackage{mathtools}
\usepackage{upgreek}
\usepackage{dsfont}

\newcommand{\doi}[1]{https://doi.org/#1}

\title{Test of validity of the force-field approximation with AMS-02 and PAMELA monthly fluxes}

\ShortTitle{Validity of the FFA}

\author{\speaker{Claudio Corti}\\
        University of Hawaii at Manoa\\
        E-mail: \email{corti@hawaii.edu}}
\author{Veronica Bindi\\
   University of Hawaii at Manoa}
\author{Cristina Consolandi\\
   University of Hawaii at Manoa}
\author{Christopher Freeman\\
   University of Hawaii at Manoa}
\author{Andrew Kuhlman\\
   University of Hawaii at Manoa}
\author{Cristopher Light\\
   University of Hawaii at Manoa}
\author{Matteo Palermo\\
   University of Hawaii at Manoa}
\author{Siqi Wang\\
   University of Hawaii at Manoa}

\abstract{Galactic cosmic rays (GCRs) entering the heliosphere are disturbed by the magnetic field of the Sun, which varies the shape and intensity of their local interstellar spectrum.
The force-field approximation is a popular way of dealing with solar modulation, especially for studies focused on galactic transport of cosmic rays.
The validity of this approach to reproduce the modulated GCR fluxes at Earth is tested using monthly proton fluxes measured by PAMELA between July 2006 and January 2014 and monthly proton and helium fluxes measured by AMS-02 between May 2011 and May 2017.
We show that the precision of the new AMS-02 data requires a rigidity-dependent modification of the force-field approximation.}

\FullConference{36th International Cosmic Ray Conference -ICRC2019-\\
		July 24th - August 1st, 2019\\
		Madison, WI, U.S.A.}

\begin{document}

\section{Introduction}
Galactic cosmic rays (GCRs) provide key information about some of the most energetic phenomena in our galaxy.
Upon entering the heliosphere, GCRs interact with the magnetic field embedded in the solar wind plasma \cite{bib:parker58:solar-wind,bib:parker65:modulation} and, as a consequence, their spectrum measured at Earth is significantly different from the spectrum in interstellar space.
The heliospheric magnetic field (HMF) changes in time according to the level of solar activity, which varies with a period of nearly 11 years, and the resulting time dependence of the GCR spectrum is called solar modulation \cite{bib:potgieter13:solar-modulation}.
\emph{Parker}, in 1965, derived the equation of transport of GCRs in the heliosphere \cite{bib:parker65:modulation}:
\begin{equation} \label{eqn:numerical-model:parker}
   \frac{\partial f}{\partial t} + \mathbf{V}_{sw} \cdot \boldsymbol{\nabla}f - \boldsymbol{\nabla} \cdot \left( \boldsymbol{\mathsf{K}} \boldsymbol{\nabla}f \right) - \frac{\boldsymbol{\nabla} \cdot \mathbf{V}_{sw}}{3} \frac{\partial f}{\partial \mathrm{ln} R} = 0,
\end{equation}

where $f(\mathbf{r},R)$ is the omni-directional GCR distribution function at position $\mathbf{r}$ and rigidity $R$, $\mathbf{V}_{sw}$ is the solar wind speed, $\boldsymbol{\mathsf{K}}$ is the diffusion tensor, containing both the diffusion and drift coefficients. 
The full three-dimensional Parker equation can not be solved analytically, and so many sophisticated numerical models have been developed.
However, these models are usually too complex to be used in the study of galactic acceleration and propagation of GCRs, and thus a much simpler description of the solar modulation process, based on the force-field approximation (FFA) \cite{bib:gleeson68:ffa}, is routinely adopted instead.
Assuming that the convective and diffusive fluxes cancel out (\emph{i.e.} zero streaming), \emph{Gleeson \& Axford} derived an analytical solution of the one-dimensional steady-state Parker equation, which relates the GCR spectrum measured at Earth, $J$, with the one in interstellar space, called local interstellar spectrum (LIS), $J_{LIS}$:
\begin{equation} \label{eqn:ffa}
   \frac{J(T_{E})}{R^{2}_{E}} = \frac{J_{LIS}(T_{HP})}{R^{2}_{HP}} \quad \Longrightarrow \quad J(T_{E}) = \frac{T_{E}(T_{E} + 2M)}{T_{HP}(T_{HP} + 2M)}J_{LIS}(T_{HP}),
\end{equation}

where $T$ is the kinetic energy, $M$ is the particle mass, and the subscripts ${}_{E}$ and ${}_{HP}$ refer, respectively, to the Earth and the heliopause, which is considered the modulation boundary.
Positing that the radial and rigidity dependence of the diffusion coefficient can be separated, $k(r,R) = \beta k_{1}(r) k_{2}(R)$, $R_{E}$ and $R_{HP}$ are related by the condition:
\begin{equation}
\int_{R_{E}}^{R_{HP}} \frac{\beta k_{2}(R)}{R} \mathrm{d}R = \int_{r_{E}}^{r_{HP}} \frac{V_{sw}(r)}{3 k_{1}(r)} \mathrm{d}r \equiv \phi(r_{E}),
\end{equation}

where $\beta = v/c$, and $\phi$ is called the modulation potential.
If $k_{2}(R) \propto R$, then the integral above yields $T_{HP} = T_{E} + Z\phi$, where $Z$ is the particle charge number and $\phi$ in theory can be computed directly from the radial dependence of the solar wind speed and diffusion coefficient.
In practice, $\phi$ is considered a free nuisance parameter while deriving the LIS from modulated GCR data.
However, it is well known that the best-fitted value of $\phi$ depends on the shape of the LIS \cite{bib:herbst10:lis-phi}, which means that fitting directly GCR data to Equation \ref{eqn:ffa} is equivalent to testing the combined hypothesis of the shape of the LIS and the assumptions of the FFA.

In this work, we derived a method to remove the dependence on the LIS, so that the validity of only the assumptions of the FFA can be checked.
We performed the test using the monthly proton fluxes measured by PAMELA between July 2006 and February 2014 \cite{bib:adriani13:pamela-p-monthly-flux,bib:martucci18:pamela-p-monthly-flux}, and the monthly proton and helium fluxes collected by AMS-02 between May 2011 and May 2017 \cite{bib:aguilar18:ams-monthly-phe}.
These datasets cover the period from the descending phase of solar cycle 23, with its unusual and long minimum occurred in 2009 and 2010, to the descending phase of solar cycle 24.

\section{LIS-independent test of the force-field approximation}
Let's consider two fluxes measured at times $t_{1}$ and $t_{2}$.
We can invert Equation \ref{eqn:ffa} applied at $t_{1}$ to express the LIS as function of $J(T,t_{1})$:
\begin{equation}
   J_{LIS}(T) = \frac{T(T+2M)}{(T-Z\phi_{1})(T-Z\phi_{1}+2M)}J(T-Z\phi_{1},\,t_{1}),
\end{equation}

and plug it back in Equation \ref{eqn:ffa} applied at $t_{2}$:
\begin{equation} \label{eqn:ffa-no-lis}
   \begin{aligned}
      J(T,\,t_{2}) & = \frac{T(T+2M)}{(T+Z\phi_{2})(T+Z\phi_{2}+2M)}J_{LIS}(T+Z\phi_{2}) \\
                   & = \frac{T(T+2M)}{(T+Z\Delta\phi)(T+Z\Delta\phi+2M)}J(T+Z\Delta\phi_{1},\,t_{1}),
   \end{aligned}
\end{equation}

where $\Delta\phi = \phi_{2} - \phi_{1}$.
Essentially, the modulated flux at $t_{2}$ is related to the modulated flux at $t_{1}$ with the same formula of the FFA, but where the modulation potential is replaced by $\Delta\phi$, while $J(T,\,t_{1})$ acts as the LIS.
This way, by choosing a given period, we can use the corresponding flux as reference flux and use it to calculate the modulated flux at different periods according to the FFA, without any assumption on the shape of the LIS.

\section{Methodology and results}

Our strategy is to perform a least-square fit of PAMELA and AMS-02 data with Equation \ref{eqn:ffa-no-lis}, where $\Delta\phi$ is the only free parameter.
We are not interested in the best-fit value of $\Delta\phi$, but in the goodness of fit as defined by the $\chi^{2}$, so that we can count how many times the fit fails.
Under the assumption that the FFA is valid, the probability for the fit to not be consistent with the data at a confidence level $s$ is $p = 1 - \mathrm{erf}(s/\sqrt{2})$, where $s$ is the number of standard deviations ($\sigma$) containing the probability corresponding to the confidence level.
Given $n$ fits, the probability of observing $k$ failed fits is distributed as a binomial with $n$ trials and event probability $p$, \emph{i.e.} we expect $pn \pm \sqrt{p(1-p)n}$ fits.

For each dataset, we chose as reference the period with the highest flux: January 2 -- 23, 2010 for PAMELA (corresponding to the solar minimum of solar cycle 23/24), and February 12 -- March 16, 2017 for AMS-02 (corresponding to the descending phase of solar cycle 24).
In order to reduce the statistical fluctuations and to be able to evaluate Equation \ref{eqn:ffa-no-lis} at any kinetic energy, the reference flux was parametrized with a 6-knot cubic spline, extrapolated with a power-law below and above the energy range of the measured fluxes.

\subsection{PAMELA protons}
\begin{figure}[h!]
   \centering
   \includegraphics[width=0.944\textwidth]{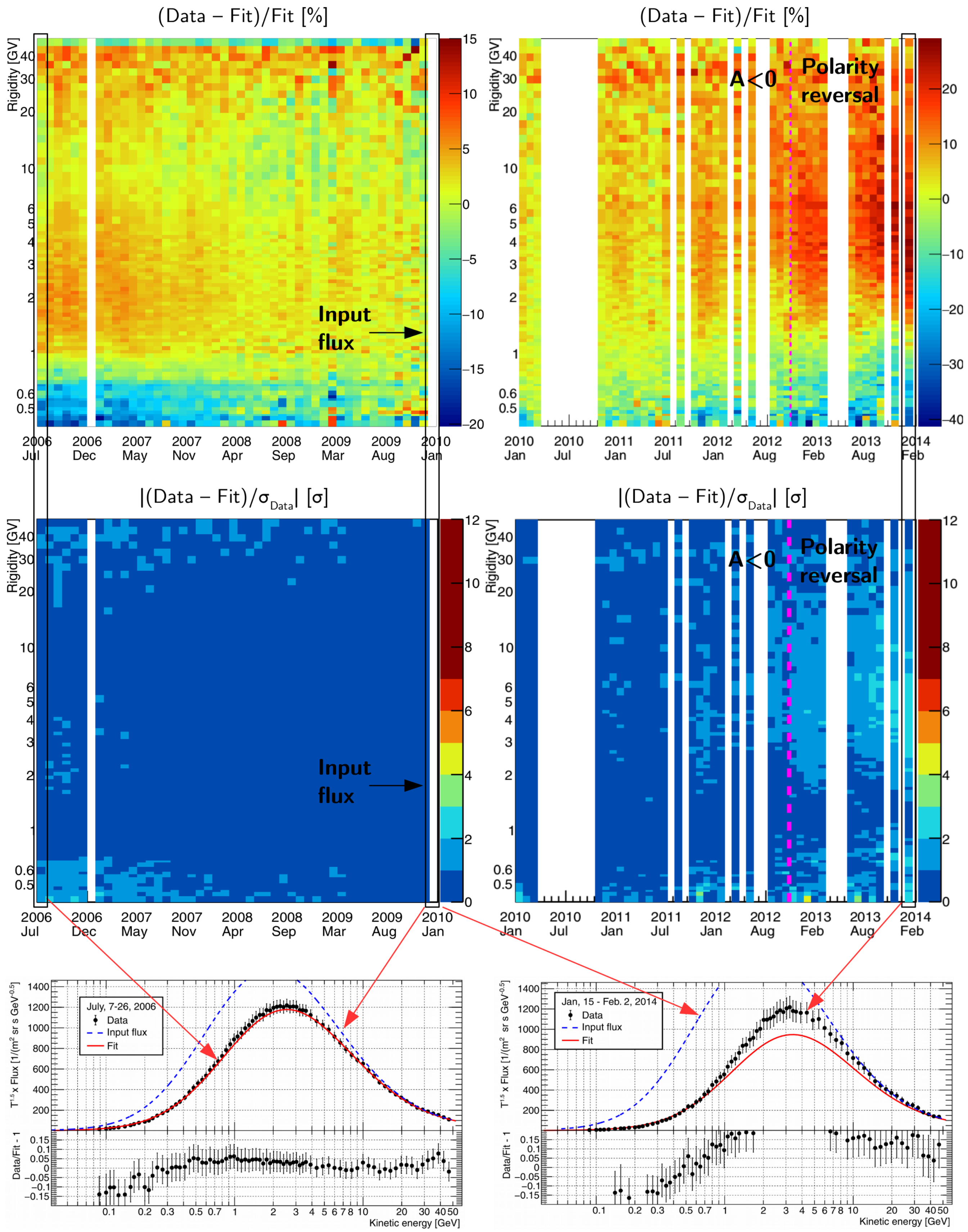}
   \caption{
      \textbf{Top.} Percent residuals (color map) of the fits to PAMELA proton data (left: Jul. 2006 -- Dec. 2009; right: Jan. 2010 -- Feb. 2014), as function of time and rigidity.
      The vertical magenta dashed line marks the beginning of the solar magnetic field polarity reversal.
      \textbf{Middle.} Same as the top figure, but for the $z$-score instead of the percent residuals.
      \textbf{Bottom.} Examples of fits (upper panel) with the corresponding relative residuals (lower panel), as function of kinetic energy, for two periods (left: Jul. 7 -- 26, 2006; right: Jan. 15 -- Feb. 2, 2014).
      The black markers are the data points (multiplied by $T^{1.5}$ to emphasize the rigidity range around 5 GV), the red solid line is the fit, the blue dashed line is the reference flux.
   }
   \label{fig:pamela}
\end{figure}

Figure \ref{fig:pamela} illustrates the results of the fits on PAMELA protons.
Data from July 2006 to December 2009 are on the left, while data from January 2010 to February 2014 are on the right.
The top part shows the percent residuals, (data - fit)/fit, as function of time and rigidity: red and blue colors correspond, respectively, to regions where the fits underestimate and overestimates the data.
The vertical magenta dashed line in the right plot marks the beginning of the solar magnetic field polarity reversal.
The middle part shows the $z$-score, |data-fit|/$\sigma_{\mathrm{Data}}$, with $\sigma_{\mathrm{Data}}$ the data uncertainty, in a similar fashion as the top part: here dark blue colors represent fits consistent with data within the error bars.
The bottom part shows two examples of fits with the corresponding residuals: the black markers are the data, the red solid line is the fit, and the blue dashed line is the reference flux.

During the descending phase of solar cycle 23, the modulation is underestimated (\emph{i.e} the fits are higher than the data) by more than $10\%$ below 0.5 GV and overestimated (\emph{i.e.} the fits are lower than the data) by $5\%$ between 1 and 5 GV with respect to the minimum of solar cycle 23/24.
During the maximum of solar cycle 24, the modulation is underestimated by more than $15\%$ below 0.5 GV and overestimated by more than $20\%$ between 2 and 20 GV with respect to the minimum of solar cycle 23/24.
Nevertheless, most of the fits are within one standard deviation from the data in all the rigidity range measured by PAMELA, except after 2013, when the discrepancies at 5 GV start to increase, reaching three standard deviations in 2014, which corresponds to the solar maximum.
Going from low to high solar activity, there is a clear trend for the modulated fluxes by the FFA becoming worse at describing the observations.

\subsection{AMS-02 protons and helium}
\begin{figure}[h!]
   \centering
   \includegraphics[width=0.96\textwidth]{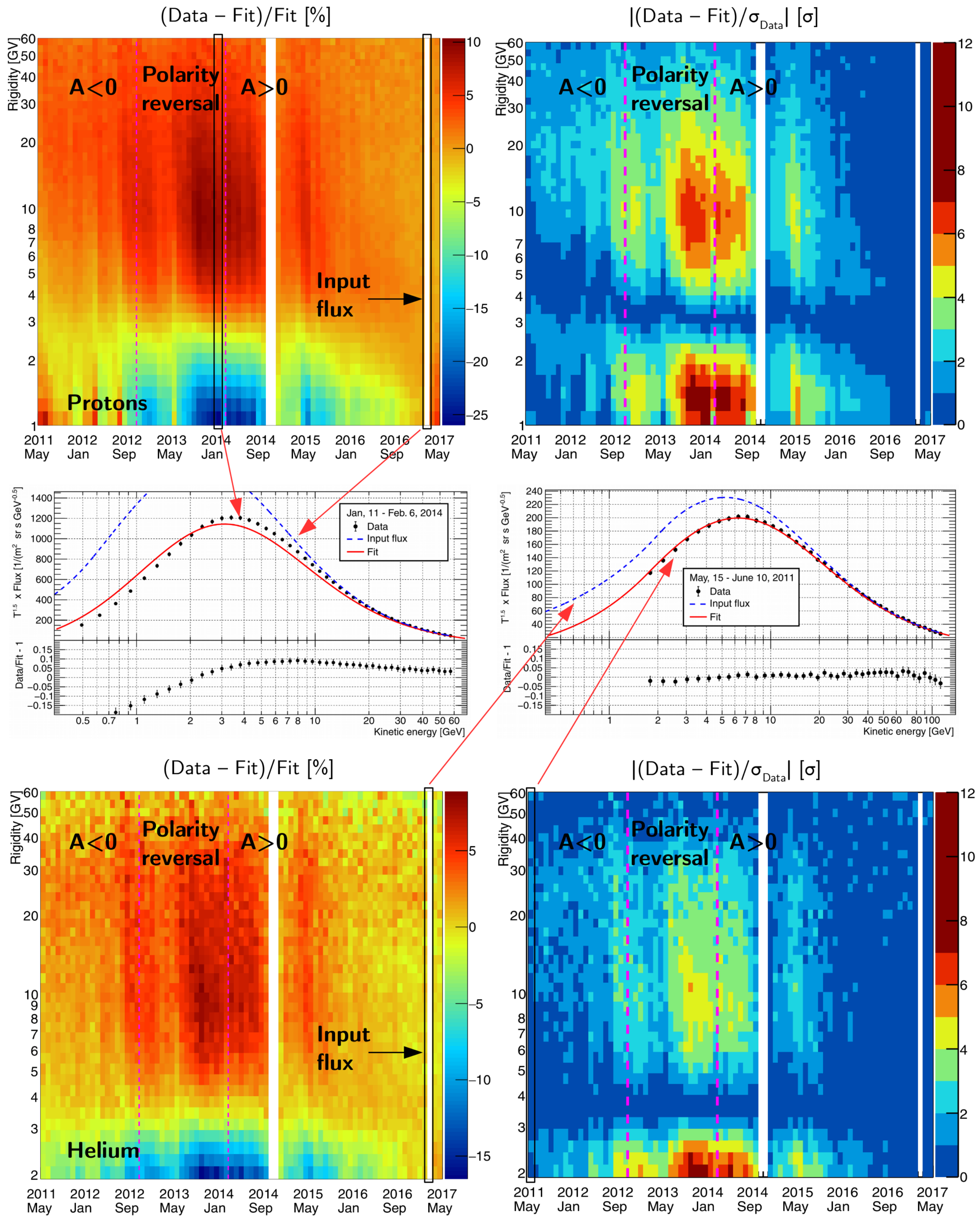}
   \caption{
      \textbf{Top.} Percent residuals (left) and $z$-score (right) of the fits to AMS-02 proton data, as function of time and rigidity.
      The vertical magenta dashed lines delimit the period of the solar magnetic field polarity reversal.
      \textbf{Middle.} Example of fits (upper panel) with the corresponding relative residuals (lower panel), as function of kinetic energy, for two periods (left: protons, Jan. 11 -- Feb. 6, 2014; right: helium, May 15 -- Jun. 10, 2011).
      The black markers are the data points (multiplied by $T^{1.5}$ to emphasize the rigidity range around 5 GV), the red solid line is the fit, the blue dashed line is the reference flux.
      \textbf{Bottom.} Same as the top figure, but for AMS-02 helium data.
   }
   \label{fig:ams}
\end{figure}

Figure \ref{fig:ams} illustrates the results of the fits on AMS-02 protons (top part) and helium (bottom part).
The top and bottom left parts show the percent residuals, similarly to Figure \ref{fig:pamela} top, while the top and bottom right parts show the $z$-score, as in Figure \ref{fig:pamela} middle.
Here, the two vertical magenta dashed lines delimit the period of the solar magnetic field polarity reversal.
The middle part shows an example of fit with the corresponding residuals for protons (left) and helium (right).

During the maximum of solar cycle 24, the modulation of protons is underestimated by more than $20\%$ below 2 GV and overestimated by more than $5\%$ between 5 and 20 GV with respect to the descending phase of solar cycle 24.
Due to the smaller uncertainties of AMS-02 protons, most of the fits are not consistent with data within the error bars, with the largest discrepancies, over six standard deviations, occurring in 2014, below 2 GV and around 7 GV.
For helium, during the maximum of solar cycle 24, the modulation is underestimated by more than $12\%$ below 2 GV and overestimated by more than $5\%$ between 5 and 20 GV with respect to the descending phase of solar cycle 24.
Also in this case, the majority of the fits are not consistent with data, with discrepancies larger than four standard deviations during the maximum, in the same rigidity ranges as protons.
Analyzing more carefully the time dependence of the residuals, we can identify three periods during which the match with the reference flux is worst: August 2012 -- February 2013, May 2013 -- September 2014, and April -- July 2015.
These dates corresponds to periods in which the GCR flux decreased suddenly and remained suppressed for months, suggesting a more global reshaping of the heliospheric conditions instead of a local solar wind transient.

\subsection{Statistical significance of FFA validity tests}
To quantify the statistical significance of the non-validity of the FFA, we can compute the $p$-value of the $\chi^{2}$ test.
For each dataset and confidence level $s$, given $n$ fluxes, $n-1$ have been fitted, so we expect $p(n-1)$ failed fits, while we observed $k$.
The significance is defined as the number of standard deviations associated to the $p$-value, 
$\sum_{m=k}^{n-1} B(m|n-1,p)$, where $B$ is the binomial probability density function.
Table \ref{tab:significance} report the number of expected and observed failed fits for various confidence levels ($s$ from $1\sigma$ to $5\sigma$), together with the significance of the non-validity of the FFA, separately for the three datasets used.
   
\begin{table}[h!]
   \setlength{\tabcolsep}{3pt}
   \begin{tabular}{c||c|c|c||c|c|c||c|c|c}
         & \multicolumn{3}{c||}{PAMELA protons} & \multicolumn{3}{c||}{AMS-02 protons} & \multicolumn{3}{c}{AMS-02 helium} \\
      \hline
      CL & Exp. & Obs. & Sign. $[\sigma]$ & Exp. & Obs. & Sign. $[\sigma]$ & Exp. & Obs. & Sign. $[\sigma]$ \\
      \hline
      $68.3\%\ (1\sigma)$    & $[22,\ 30]$ & 11 &   0 & $[21,\ 29]$ & 61 &  8.4 & $[21,\ 29]$ & 50 &  5.8 \\
      $95.4\%\ (2\sigma)$    & $[2,\ 6]$   & 10 & 2.9 & $[2,\ 5]$   & 59 & 16.7 & $[2,\ 5]$   & 48 & 13.9 \\
      $99.7\%\ (3\sigma)$    & $[0,\ 1]$   &  6 & 5.3 & $[0,\ 1]$   & 56 & 23.8 & $[0,\ 1]$   & 41 & 19.4 \\
      $99.99\%\ (4\sigma)$   & 0 			 &  4 & 6.7 & 0			  & 52 & 30.1 & 0				 & 39 & 25.4 \\
      $99.9999\%\ (5\sigma)$ & 0 			 &  2 & 6.1 & 0			  & 49 & 36.1 & 0			 	 & 35 & 29.9 \\
   \end{tabular}
   \caption{
      Number of expected and observed failed fits, and statistical significance of rejecting the FFA at different confidence levels, for the datasets used in this work.
   }
   \label{tab:significance}
\end{table}

For AMS-02, the significance is always greater than $5\sigma$ at all confidence levels.
For PAMELA, although the residuals seemed to indicate an overall good agreement between the fits and the data, the number of fluxes not consistent with data within three or more standard deviations is too large, and the significance of rejecting the modulation predicted from the FFA is more than $5\sigma$.
The uncertainty reported by experiments is usually understood representing the standard deviation of a normal distribution centered on the observed quantity, but we acknowledge that extrapolating this assumption to the tails of the uncertainty distribution is probably not correct, so that the significance obtained for $s \geqslant 3$ is most probably overestimated.
Even so, the significance for $s=2$ is almost $3\sigma$.
Furthermore, the failed fits on the PAMELA dataset are all clustered after 2013, during the ascending phase and solar maximum of solar cycle 24, consistent with the AMS-02 results.
%


\section{Conclusions}

In this work, we devised a method to test the validity of the FFA, independently from any LIS.
Using monthly GCR fluxes measured by PAMELA and AMS-02 between July 2006 and May 2017 we found that the solar modulation behaves differently between periods with low and high solar activity.
The modulated fluxes predicted by the FFA are not consistent with the observations with a significance over $5\sigma$.
In particular, the largest discrepancies are observed below 2 GV, where the modulation can be underestimated by up to $20\%$, and around $5 \div 10$ GV, where the modulation can be overestimated by more than $10\%$.

The key assumptions used in the FFA derivation are: a steady-state system, spherical symmetry, zero streaming, and a diffusion coefficient proportional to $R$.
The steady-state approximation is obviously less valid during periods of high solar activity, when the heliosphere is constantly changing.
Nevertheless, two- and three-dimensional steady-state numerical models have been able to reproduce AMS-02 data during the solar maximum \cite{bib:tomassetti18:model,bib:corti19:model}.
The spherical symmetry approximation removes any effect of particle drifts, as the drift velocity field, particularly along the current sheet, is inherently a three-dimensional process.
AMS-02 collected data during the polarity reversal of the solar magnetic field, observing a clear charge-sign dependence in the modulation of electrons and positrons \cite{bib:aguilar18:ams-monthly-elepos}, but the proton and helium data used here are not sufficient to draw any conclusion about the effect of neglecting drifts on the validity of the FFA.
The zero streaming hypothesis is valid in the inner heliosphere only above 400 MeV/n \cite{bib:gleeson68:ffa}, so this could explain the discrepancies below 1 GV observed by PAMELA.
On the other hand, turbulence theory shows that assuming $k_{2} \propto R$ is a crude approximation, as the rigidity dependence of the diffusion coefficient is expected to flatten below few GV due to the transition from the energy-injection to the inertial range in the HMF power spectrum \cite{bib:jokipii71:modulation}.
Indeed, \emph{Corti et al.} \cite{bib:corti16:lis-modffa} and \emph{Gieseler et al.} \cite{bib:gieseler17:modffa} proposed to modify the FFA by introducing a rigidity-dependent modulation potential which effectively reproduce the effect of a change in the rigidity dependence of the diffusion coefficient.

\acknowledgments
This work has been supported by: National Science Foundation Early Career under grant (NSF AGS-1455202); Wyle Laboratories, Inc. under grant (NAS 9-02078); NASA under grant (17-SDMSS17-0012).


\begin{thebibliography}{99}
   \bibitem{bib:adriani13:pamela-p-monthly-flux}
   O. Adriani et al. (PAMELA collaboration),
   \emph{Time dependence of the proton flux measured by PAMELA during the 2006 July--2009 December solar minimum},
   \href{\doi{10.1088/0004-637X/765/2/91}}{\emph{ApJ} {\bf 765} (2013) 91}.
   
   \bibitem{bib:aguilar18:ams-monthly-phe}
   M. Aguilar et al. (AMS-02 collaboration),
   \emph{Observation of fine time structures in the cosmic proton and helium fluxes with the Alpha Magnetic Spectrometer on the International Space Station},
   \href{\doi{10.1103/PhysRevLett.121.051101}}{\emph{PhRvL} {\bf 121} (2018) 051101}.
   
   \bibitem{bib:aguilar18:ams-monthly-elepos}
   M. Aguilar et al. (AMS-02 collaboration),
   \emph{Observation of complex time structures in the cosmic-ray electron and positron fluxes with the Alpha Magnetic Spectrometer on the International Space Station},
   \href{\doi{10.1103/PhysRevLett.121.051102}}{\emph{PhRvL} {\bf 121} (2018) 051102}.
   
   
   \bibitem{bib:corti16:lis-modffa}
   C. Corti et al.,
   \emph{Solar modulation of the local interstellar spectrum with Voyager 1, AMS-02, PAMELA and BESS},
   \href{\doi{10.3847/0004-637X/829/1/8}}{\emph{ApJ} {\bf 829} (2016) 8}.
   
   \bibitem{bib:corti19:model}
   C. Corti et al.,
   \emph{Numerical modeling of galactic cosmic-ray proton and helium observed by AMS-02 during the solar maximum of solar cycle 24},
   \href{\doi{10.3847/1538-4357/aafac4}}{\emph{ApJ} {\bf 871} (2019) 253}.

   \bibitem{bib:gieseler17:modffa}
   J. Gieseler et al.,
   \emph{An empirical modification of the force field approach to describe the modulation of galactic cosmic rays close to earth in a broad range of rigidities},
   \href{\doi{10.1002/2017JA024763}}{\emph{JGRA} {\bf 122} (2017) 10964}.
   
   \bibitem{bib:gleeson68:ffa}
   L.J. Gleeson \& W.I. Axford,
   \emph{Solar modulation of galactic cosmic rays},
   \href{\doi{10.1086/149822}}{\emph{ApJ} {\bf 154} (1968) 1011}.
   
   \bibitem{bib:herbst10:lis-phi}
   K. Herbst et al.,
   \emph{On the importance of the local interstellar spectrum for the solar modulation parameter},
   \href{\doi{10.1029/2009JD012557}}{\emph{JGRD} {\bf 115} (2010) 00I20}.
   

   \bibitem{bib:jokipii71:modulation}
   R.J. Jokipii,
   \emph{Propagation of cosmic rays in solar wind},
   \href{\doi{10.1029/RG009i001p00027}}{\emph{RvGSP} {\bf 9} (1971) 27}.

   \bibitem{bib:martucci18:pamela-p-monthly-flux}
   M. Martucci et al. (PAMELA collaboration),
   \emph{Proton fluxes measured by the PAMELA experiment from the minimum to the maximum solar activity for solar cycle 24},
   \href{\doi{10.3847/2041-8213/aaa9b2}}{\emph{ApJ} {\bf 854} (2018) L2}.
   
   \bibitem{bib:parker58:solar-wind}
   E.N. Parker,
   \emph{Dynamics of the interplanetary gas and magnetic fields},
   \href{\doi{10.1086/146579}}{\emph{ApJ} {\bf 128} (1958), 664}.
   
   \bibitem{bib:parker65:modulation}
   E.N. Parker,
   \emph{The passage of energetic charged particles through interplanetary space},
   \href{\doi{10.1016/0032-0633(65)90131-5}}{\emph{P\&SS} {\bf 13} (1965) 9}.
   
   \bibitem{bib:potgieter13:solar-modulation}
   M. Potgieter,
   \emph{Solar modulation of cosmic rays},
   \href{\doi{10.12942/lrsp-2013-3}}{\emph{LRSP} {\bf 10} (2013) 3}.


   \bibitem{bib:tomassetti18:model}
   N. Tomassetti et al.,
   \emph{Testing diffusion of cosmic rays in the heliosphere with proton and helium data from AMS},
   \href{\doi{10.1103/PhysRevLett.121.251104}}{\emph{PhRvL} {\bf 121} (2018) 251104}.

   
\end{thebibliography}
\end{document}